\documentclass[copyright,creativecommons,noderivs,noncommercial]{eptcs}
\providecommand{\event}{SSV 2012}
\usepackage{breakurl}
\usepackage{hyperref}
\usepackage[utf8]{inputenc}

\usepackage{gastex}
\usepackage{amsmath}
\usepackage{amsfonts}
\usepackage{amssymb}

\usepackage{ifpdf}

\ifpdf
\usepackage[pdftex]{graphicx}
\else
\usepackage[dvips]{graphicx}
\fi

\usepackage[rflt]{floatflt}

\usepackage{listings}

\usepackage{datatool}
\usepackage{intcalc}
\usepackage{numprint}
\usepackage{rotating}
\usepackage{xstring}
\usepackage{subfigure}
\usepackage{caption}

\usepackage{url}

\lstdefinestyle{C}{language=C,
  basicstyle=\fontsize{8}{10}\selectfont\sffamily,
  columns=fullflexible
}

\lstdefinestyle{C-inline}{language=C,
  basicstyle=\sffamily,
  columns=fullflexible
}

\newcommand{\headC}[1]{\multicolumn{1}{c|}{#1}}

\newcommand{\tudparagraph}[1]{%

\vspace{.5ex}
\noindent%
\textbf{#1}%
}

\begin{document}

\title{Chiefly Symmetric: Results on the Scalability of Probabilistic Model Checking for Operating-System Code
\thanks{
  This work was in part funded by 
  the German Research Council (DFG) through the QuaOS project, 
  the collaborative research center 912 Highly-Adaptive
  Energy-Efficient Computing (HAEC),
  the European Union Seventh Framework Programme under grant
  agreement no. 295261 (MEALS),
  the DFG/NWO-project ROCKS,
  the cluster of excellence cfAED (center for Advancing
  Electronics Dresden) 
  and
  by the European
  Union and the state Saxony through the ESF young researcher
  group IMData 100098198.
}}

\author{Christel Baier$^1$, Marcus Daum$^1$, Benjamin Engel$^2$, 
        Hermann H\"artig$^2$, Joachim Klein$^1$ \\
        Sascha Kl\"uppelholz$^1$, Steffen M\"arcker$^1$,
        Hendrik Tews$^2$, Marcus V\"olp$^2$
\institute{%
$^1$Institute for Theoretical Computer Science and
$^2$Operating-Systems Group \\
Technische Universit\"at Dresden, Germany}
}

\def\authorrunning{C. Baier, H. H\"artig et.~al.}
\def\titlerunning{Chiefly symmetric: Scalability of Probabilistic Model Checking for OS Code.}

\maketitle


\begin{abstract}
Reliability in terms of functional properties from the safety-liveness
spectrum is an indispensable requirement of low-level operating-system
(OS) code. However, with evermore complex and thus less predictable
hardware, quantitative and probabilistic guarantees become more and
more important. Probabilistic model checking is one technique to
automatically obtain these guarantees.  First experiences with the
automated quantitative analysis of low-level operating-system code
confirm the expectation that the naive probabilistic model checking
approach rapidly reaches its limits when increasing the numbers of
processes.  This paper reports on our work-in-progress to tackle the
state explosion problem for low-level OS-code caused by the
exponential blow-up of the model size when the number of processes
grows. We studied the symmetry reduction approach and
carried out our experiments with a simple test-and-test-and-set lock
case study as a representative example for a wide range of protocols
with natural inter-process dependencies and long-run properties.  We
quickly see a state-space explosion for scenarios where inter-process
dependencies are insignificant. However, once inter-process
dependencies dominate the picture models with hundred and more
processes can be constructed and analysed.
\end{abstract}


\section{Introduction}
\label{sec:introduction}

Safety-critical software in space, flight, and automotive control
systems need not only produce correct results. It must produce these
results in time and, to the extent possible, despite component failures. Worst-case
execution-time (WCET) analyses (cf.,~\cite{BCP02,KP07,WEE+08}) are
able to produce the former kind of guarantees whereas reliability
techniques (cf.,~\cite{nooks,nonstop,Doebel12Watchmen}) are designed to rule out negative
effects from the latter. However, WCET analyses only produce
guarantees in the form of upper bounds on the execution times of all
involved components, which hold even in the most extreme
situations. Yet, most computer systems are either not safety critical
or they include fail safe mechanisms that prevent damage in highly
exceptional situations.
For such systems, it is much more appropriate to look at the
$99.9\%$ quantile of the execution time (i.e., the execution time
that is not exceeded with a probability of $99.9\%$), ignoring
exceptional cases, that are dealt with by other means anyway.

Many software systems, especially operating systems, contain
optimisations that are geared towards the average case, incurring
additional costs under exceptional circumstances. To justify such
optimisations one has to look at the probability at which the
optimisation is advantageous as well as the additional costs that
apply with a (hopefully) low probability. Here it is again often
more appropriate to investigate a suitable quantile of the
additional costs, because pathological cases are either not
interesting or solved by other means.

In previous work~\cite{FMICS12}, we presented a combined
simulation and model checking based approach to determine the
probabilistic quantitative properties similar to what we
described in the two preceding paragraphs. 
As a part of this first case study we have extended the model checker
PRISM~\cite{KwiNorPar04} with an operator for steady-state dependent
properties and performed an analysis of the long run behaviour of a
simple test-and-test-and-set 
spinlock~\cite{anderson:spin_lock_alternatives}. However, scaling the
number of processes competing for the lock proved difficult, 
as the state space to be considered by the model checker rapidly increases. 

In this paper, we report on our experience in applying symmetry
reduction~\cite{ClarEndFilJha96,EmSis96,IpDill96,EmTref99}
to mitigate the state space explosion and thus scale our
spinlock analysis. As our model is highly symmetrical, symmetry
reduction techniques allow the analysis to be carried out on a
potentially much smaller quotient model instead of the full model. 
To evaluate the benefit of applying symmetry reduction to our model
and to gain further insight about its behaviour for an
increasing number of processes, we have implemented a specialised 
tool that explores the reachable state space of 
the symmetry reduced quotient model and generates 
the transition matrix as input for a probabilistic model checker 
to carry out its analysis. This approach yielded a drastic improvement
in the scalability of our analysis, in some cases allowing an analysis
of models with 10,000 processes and more in contrast to the 
4 to 5 processes achievable with the non-reduced model. 

\tudparagraph{Outline.} In the next section, we introduce our general
approach, present our model and demonstrate the scalability problems
of the non-reduced model. Then, in Section~\ref{sec:symmetry} we
discuss the symmetry reduction techniques that we have applied. We
interpret the drastic improvements in scalability and the results of
the analysis for an increasing number of
processes. Section~\ref{sec:conclusions} concludes.


\section{The QuaOS Approach for Probabilistic Quantitative Properties}
\label{sec:quaos-approach}

In this section, we first give a short overview of the general
approach that we use in the QuaOS project for determining
probabilistic quantitative properties of operating-system code. Then,
we introduce our cache-aware model of the test-and-test-and-set
spinlock.

One very difficult task in modelling the quantitative behaviour of
low-level operating-system code is to design the necessary
abstractions such that the model agrees in its behaviour with
reality. Our approach is to compare the model checking results with
measurements from specifically designed test cases. In case there are
deviations, we investigate the reasons and adjust the model and, if
necessary, the test cases to extract further
details. For~\cite{FMICS12} we found that a specific effect of
cache coherence has a substantial influence on the timing
of spinlocks.

Our ultimate aim is to make predictions for hardware
configurations that do not yet exist. With our approach, we could
for instance investigate which synchronisation primitives are
suitable for systems with more than 100 cores, depending on the
work load and the performance of certain memory operations.


%
%
\tudparagraph{A probabilistic cache-aware model of
  test-and-test-and-set spinlocks.}  Figure~\ref{fig:spinlock} shows the
source code for a test-and-test-and-set
lock~\cite{anderson:spin_lock_alternatives}.  In the following, we
refer to this lock simply as spinlock. The \textsf{atomic\_swap} in
line 3 writes the second argument into the location of the first
argument and returns the old value of the first argument.  Therefore,
the lock has been acquired, if \textsf{atomic\_swap} returns
\textsf{false}. Otherwise, the caller spins in line 4 until the lock
becomes free before it executes the more expensive
\textsf{atomic\_swap} again. Note that each access of the shared
variable \textsf{occupied} might cause the transfer of the \newpage

\begin{floatingfigure}{.51\textwidth}
  \vspace{-2mm}
  \begin{center}
 \begin{tabular}{r@{\hspace*{0.3cm}}l}
   \\[-4ex]
   {\footnotesize 1} & {\sf volatile} bool occupied = false;
   \\[1ex]
   {\footnotesize 2} & {\sf void} lock()\{
   \\
   {\footnotesize 3} & \quad {\sf while}(atomic\_swap(occupied,true))\{
   \\
   {\footnotesize 4} & \qquad {\sf while}(occupied)\{\}
   \\
   {\footnotesize 5} & \quad \} \}
   \\[1ex]
   {\footnotesize 6} & {\sf void} unlock()\{
   \\
   {\footnotesize 7} & \quad occupied = false
   \\
   {\footnotesize 8} & \}
 \end{tabular}
 \vspace{-1ex}
  \caption{Pseudo Code for a TTS spinlock}
  \label{fig:spinlock}
 \vspace{-1ex}
  \end{center}
\end{floatingfigure}

\noindent
respective cache line into the core-local cache.

In our investigation we consider $n$ identical parallel processes $P_i$
that use the spinlock to synchronise their critical sections.
Each process $P_i$ performs an uncritical section and a
critical section in an endless loop. The durations of the
uncritical and the critical sections are determined by
probabilistic distributions, which are parameters in our model.

\begin{figure}[tbhp]
  \begin{center}
    \small
    \setlength{\unitlength}{2.0pt}
    \begin{picture}(224,70)(-32,-64)

\gasset{AHLength=3}

\node[Nadjust=wh,Nmr=5](start)(15,-7){%
	\begin{tabular}{c}
		 $\mathit{start}_i$
	\end{tabular}}
 
\node[Nadjust=wh,Nmr=5](nc)(80,-7){%
	\begin{tabular}{c}
		 $\mathit{ncrit}_i$
	\end{tabular}}

\node[Nadjust=wh,Nmr=5](wait)(145,-60){%
	\begin{tabular}{c}
		 $\mathit{wait}_i$
	\end{tabular}}
	 
\node[Nadjust=wh,Nmr=5](crit)(15,-60){%
	\begin{tabular}{c}
		$\mathit{crit}_i$
	\end{tabular}}

\imark[ilength=12,iangle=130](start)

\drawedge[ELside=l,ELpos=50](start,nc){%
	 \begin{tabular}{c}
		$\mathit{initialize}$: \\[-0.5ex]
		$t_i := \mathit{random}(\nu)$
	 \end{tabular}}			

\drawedge[ELside=l,ELpos=60,curvedepth=-4](nc,wait){%
	 \begin{tabular}{c}
		if $t_i {=}0$ then $\mathit{tick}$ 
	 \end{tabular}}			

\drawedge[ELside=r,ELpos=50](wait,crit){%
	 \begin{tabular}{c}
		if $\mathit{lock}_i \land t_i{=}1$ then $\mathit{tick}$: 
                $t_i := \mathit{random}(\gamma_0)$
                \\
		if $\mathit{lock}_i \land t_i{=}2$ then $\mathit{tick}$: 
                $t_i := \mathit{random}(\gamma_1)$
	\end{tabular}}

\drawedge[ELside=l,ELpos=45,curvedepth=-4](crit,nc){%
	 \begin{tabular}{c}
		if $t_i{=}0$ then $\mathit{tick}$: \\[-0.5ex]
		$t_i := \mathit{random}(\nu)$
	 \end{tabular}}			

\drawloop[loopdiam=17,loopangle=0,ELpos=50](nc){%
	 \begin{tabular}{c}
		if $t_i > 0$ then $\mathit{tick}$: \\[-0.5ex]
		$t_i := t_i{-}1$
	 \end{tabular}}

\drawloop[loopdiam=17,loopangle=90,ELpos=77](wait){%
	 \begin{tabular}{c}
		if $\neg \mathit{lock}_i$ 
		then $\mathit{tick}$: \\[-0.5ex]
		$t_i := \min\{t_i{+}1,2\}$
	 \end{tabular}}

\drawloop[loopdiam=17,loopangle=90,ELpos=23](crit){%
	 \begin{tabular}{c}
		if $t_i > 0$ then $\mathit{tick}$: \\[-0.5ex]
		$t_i := t_i{-}1$
	 \end{tabular}}

\end{picture}

  \end{center}
  \caption{Control flow graph of process $P_i$
    \label{fig:control-flow-proc}}
\end{figure}
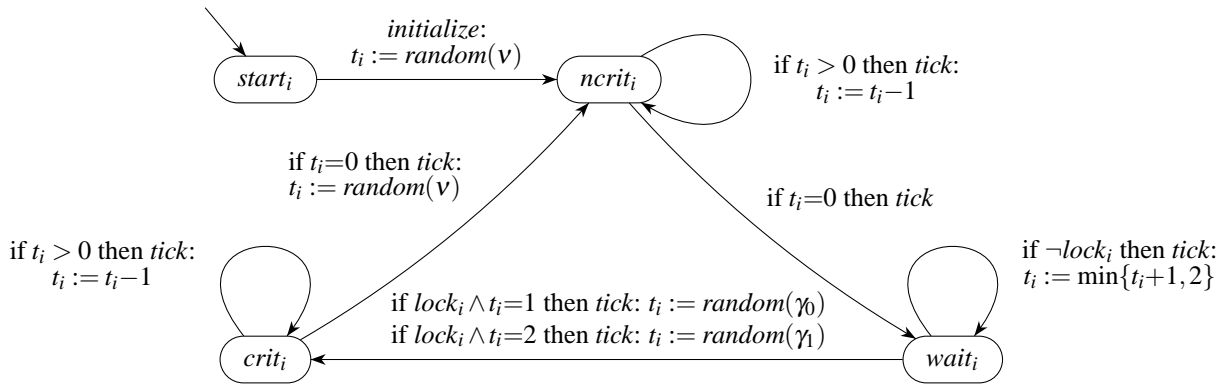

\begin{figure}[tbp]
  \begin{center}
    \small
    \setlength{\unitlength}{2.0pt}
    \begin{picture}(222,86)(-36,-85)

\gasset{AHLength=3}

\node[Nadjust=wh,Nmr=5](unlock)(75,-20){%
	 \begin{tabular}{c}
		 $\mathit{unlock}$ 
	 \end{tabular}}

\node[Nadjust=wh,Nmr=5](lock_i)(20,-70){%
	 \begin{tabular}{c}
		 $\mathit{lock}_i$ 
	 \end{tabular}}

\node[Nadjust=wh,Nframe=n](dots1)(0,-70){%
	 \begin{tabular}{c}
		 $\ldots$
	 \end{tabular}}

\node[Nadjust=wh,Nframe=n](dots2)(150,-70){%
	 \begin{tabular}{c}
		 $\ldots$
	 \end{tabular}}

\node[Nadjust=wh,Nmr=5](lock_k)(130,-70){%
	 \begin{tabular}{c}
		 $\mathit{lock}_k$ 
	 \end{tabular}}

\imark[ilength=18,iangle=160](unlock)

\drawedge[ELside=l,ELpos=40](unlock,lock_i){%
	 \begin{tabular}{r}
		 if $\mathit{wait}_i$ \\[-0.5ex]
		 then $\mathit{tick}$
	 \end{tabular}}

\drawedge[ELside=l,ELpos=34,curvedepth=9](lock_i,unlock){%
    \begin{tabular}{c}
	if $\mathit{crit}_i \wedge t_i{=}0$ and \\[-0.5ex]
	   $\neg\mathit{wait}_1 \wedge \ldots \wedge 
            \neg\mathit{wait}_n$\\[-0.5ex]
	then $\mathit{tick}$ 
    \end{tabular}%
    \hspace*{-0.5cm}}

\drawloop[ELside=l,ELpos=36,loopdiam=17,loopangle=135](lock_i){%
	 \begin{tabular}{c}
	   if $\neg(\mathit{crit}_i \wedge t_i{=}0)$ \\[-0.5ex]
	   then $\mathit{tick}$ 
	 \end{tabular}}

\drawedge[ELside=r,ELpos=40](unlock,lock_k){%
	 \begin{tabular}{l}
		 if $\mathit{wait}_k$ \\[-0.5ex]
		 then $\mathit{tick}$
	 \end{tabular}}

\drawedge[ELside=r,ELpos=34,curvedepth=-9](lock_k,unlock){%
   \hspace*{-0.5cm}
   \begin{tabular}{c}
	 if $\mathit{crit}_k \wedge t_k{=}0$ and \\[-0.5ex]
	    $\neg\mathit{wait}_1 \wedge \ldots \wedge 
              \neg\mathit{wait}_n$ \\[-0.5ex]
	 then $\mathit{tick}$ 
    \end{tabular}}

\drawloop[ELside=l,ELpos=64,loopdiam=17,loopangle=45](lock_k){%
	 \begin{tabular}{c}
	    if $\neg(\mathit{crit}_k \wedge t_k{=}0)$ \\[-0.5ex]
	    then $\mathit{tick}$ 
	 \end{tabular}}

\drawedge(lock_i,lock_k){%
    \begin{tabular}{l}
	if $\mathit{crit}_i \wedge t_i{=}0 \wedge \mathit{wait}_k$
	then $\mathit{tick}$ 
    \end{tabular}}

\drawedge[curvedepth=9](lock_k,lock_i){%
    \begin{tabular}{l}
	if $\mathit{crit}_k \wedge t_k{=}0 \wedge \mathit{wait}_i$
	then $\mathit{tick}$ 
    \end{tabular}}

\drawloop[ELside=l,ELpos=80,loopdiam=17,loopangle=90](unlock){%
   \begin{tabular}{c}
     if $\neg \mathit{wait}_1 \wedge 
         \ldots \wedge \neg \mathit{wait}_n$ \\[-0.5ex]
     then $\mathit{tick}$ 
   \end{tabular}}

\end{picture}

  \end{center}
  \caption{Control flow graph of the spinlock
    \label{fig:control-flow-lock}}
\end{figure}
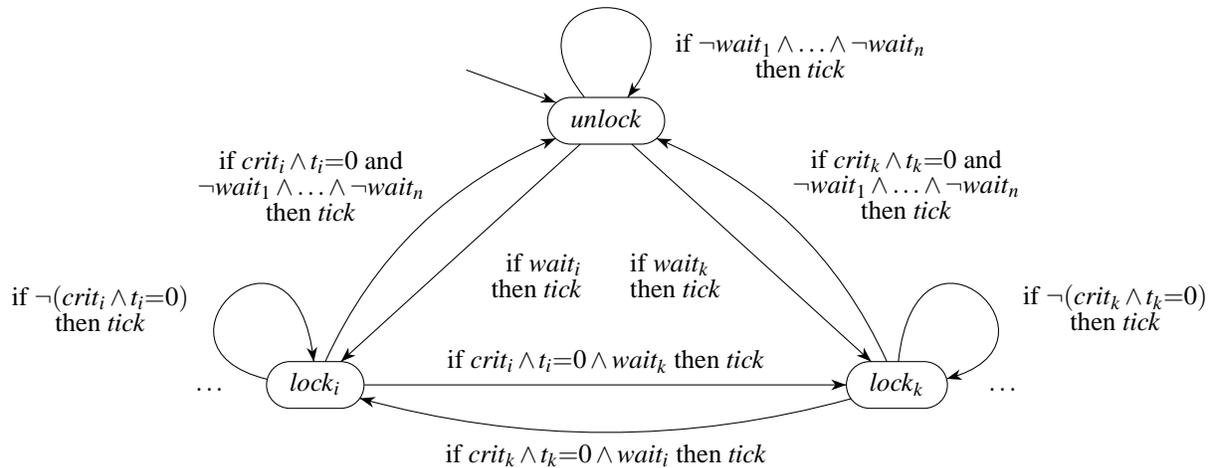

In~\cite{FMICS12} we developed a discrete time Markov chain
(DTMC) model that results from the synchronous parallel composition of
$n$ processes $P_i$ (Figure~\ref{fig:control-flow-proc}) and the
spinlock itself (Figure~\ref{fig:control-flow-lock}). Having a
separate process for the spinlock facilitates the uniform
probabilistic choice of the process that acquires the lock next, in
case several processes are spinning.

The control flow graph for each process $P_i$ (see
Figure~\ref{fig:control-flow-proc}) has the location $\mathit{ncrit}$
for the non-critical section, $\mathit{wait}$ for the time it waits
for the lock, $\mathit{crit}$ for the critical section and
$\mathit{start}$ for initialisation. In the locations $\mathit{ncrit}$
and $\mathit{crit}$ the variable $t_i$ is a timer that determines how
long the process has to remain in the respective location.  The time
for the non-critical section is determined by the distribution
$\nu$. The time for the critical section length is determined by the
distributions $\gamma_0$ and $\gamma_1$. In the location
$\mathit{wait}$ the variable $t_i$ records whether the lock could be
entered immediately ($t_i = 1$) or whether spinning was necessary
before obtaining the lock ($t_i = 2$).

We need the two distributions $\gamma_0$ and $\gamma_1$ to account for
the following cache effect. In the usual case where the last lock
owner was a different process (on a different core), the
\textsf{atomic\_swap} in Line 3 (see Figure~\ref{fig:spinlock})
transfers the cache line containing the variable \textsf{occupied}
into the local cache with state \emph{modified}. In case the process
was spinning when the lock was released, Line 4 causes an additional
inter-core message that transfers the respective cache line into state
\emph{shared}. This additional inter-core message delays processes
that were spinning before acquiring the lock by about 200 cycles. For
simplicity we model this delay by using the distribution $\gamma_1$
for the critical section, which offsets $\gamma_0$ by this delay. For
small critical sections (in the order of 1000 cycles) the delay
significantly changes the behaviour of the complete system. The values
for the distribution $\nu$ are chosen to be at least one order of
magnitude larger than $\gamma_0$. Smaller ratios between critical and
uncritical are uncommon in reality. In~\cite{FMICS12} we
compare our results with a model that uses only one distribution for
the critical section.

To reduce state-space size, we scale down
the numerical values of all distributions as far as possible while
maintaining their ratios. The main example in this paper will be
parametrised with the distributions $\gamma_0(5) = 1$, $\gamma_1(6) =
1$ and $\nu(40) = \nu(50) = \frac{1}{2}$.

The interaction of $P_i$ with the spinlock (see
Figure~\ref{fig:control-flow-lock}) is a bit subtle. When $P_i$
moves into state $\mathit{wait}_i$ it signals the spinlock its
desire to acquire the lock. In case the lock is free (state
$\mathit{unlock}$) the spinlock moves \emph{in the next step} 
to state $\mathit{lock}_i$, while $P_i$ stays in $\mathit{wait}_i$.
Having acquired the lock, $P_i$ moves into state $\mathit{crit}_i$ one step later.
The spinlock has additional transitions between all the $n$
$\mathit{lock}$ states to permit a direct change from
$\mathit{lock}_i$ to $\mathit{lock}_k$.

In~\cite{FMICS12} we have formalised four conditional long run
properties, such as the probability to acquire the lock without
waiting, the expected average waiting time, and the $95\%$ quantile of
this waiting time, and computed their results with the model checker
PRISM.  With PRISM, we were able to obtain results for all the
considered distributions only for up to 4 processes, with 5 processes
results for the simpler distributions were achievable. For more processes, PRISM did
not terminate within several days.

\begin{table}[bp]
\centering
 
\begin{tabular}{|r|r|r|r|r|} \hline
 & & & \headC{Time for} & \headC{Time for calculating} \\
Processes & 
  \headC{Reachable states} & 
  \headC{MTBDD size} & 
  \headC{building} & 
  \headC{steady state probabilities} \\ \hline
3 & 
 \npthousandsep{,}\numprint{67001} &
 \npthousandsep{,}\numprint{32772} & 
 \npthousandsep{,}\npdecimalsign{.}\nprounddigits{2}\numprint[s]{4.05} & 
 \npthousandsep{,}\npdecimalsign{.}\nprounddigits{2}\numprint[s]{3.45} \\ \hline
4 & 
 \npthousandsep{,}\numprint{4082808} &
 \npthousandsep{,}\numprint{569042} & 
 \npthousandsep{,}\npdecimalsign{.}\nprounddigits{2}\numprint[s]{316.31} & 
 \npthousandsep{,}\npdecimalsign{.}\nprounddigits{2}\numprint[s]{149.10}\\ \hline
5 & 
 \npthousandsep{,}\numprint{198808720} &  
 \npthousandsep{,}\numprint{7632342} & 
  $\approx$ 20 hours & 
  $\approx$ 90 minutes \\ \hline
\end{tabular}

 \caption{Statistics for building the model 
   (with $\gamma_0(5) = 1$, $\gamma_1(6) = 1$ and 
    $\nu(40) = \nu(50) = \frac{1}{2}$) 
   and calculating the steady state
   probabilities for $n$ processes 
   using the model checker PRISM with
   the sparse engine. The MTBDD size column lists the number of
   nodes used to store the symbolic transition relation in PRISM.
 }
\label{tab:prism-40-50}
\end{table}

Table~\ref{tab:prism-40-50} summarises the
statistics for the model with our example distributions.
All calculations in the paper were carried out using a
dual-socket Intel Xeon L5630 (Quad-Core) machine at 2.13 GHz with
192 GB total amount of RAM.
One can clearly see that increasing the number of processes in our
original model significantly increases the complexity, resulting in a
very fast blowup in the number of reachable states as well as in the
time spent analysing the model. 
In this paper we mainly focus on the
two factors -- building the model and calculating the steady state
probabilities -- that dominate the analysis time of the investigated
conditional long run properties. We will now detail our findings in
applying symmetry reduction as a method to scale the analysis.


\section{Applying Symmetry Reduction to the Spinlock Model}
\label{sec:symmetry}

Our model exhibits a high amount of symmetry.
For example, the control flow graph for each of
the $n$ processes (Figure~\ref{fig:control-flow-proc}) arises from a
simple renaming of the processes (by their index $i$). 
Furthermore, the control flow graph for the spinlock 
(Figure~\ref{fig:control-flow-lock})
treats each
process identically, i.e., there is a uniform choice between the
spinning processes to determine which particular process is awarded
the lock. We have exploited this symmetry at a basic level already
in~\cite{FMICS12} by concentrating on a particular process 
($P_1$) for the properties we analysed. Due to the symmetry between
the processes, the results will be identical for the other processes.

For our model, symmetry reduction is thus a natural candidate to
approach the state space explosion problem because 
the $n$ processes that compete for the spinlock behave identically.
The potential of
symmetry reduction for reducing the state space has been extensively
researched in the literature in non-probabilistic as well as in
probabilistic settings, see
e.g.~\cite{ClarEndFilJha96,EmSis96,IpDill96,EmTref99,KNP06,MilDonCal06,DM06}.
The basic idea is to perform the analysis on a smaller quotient model
that arises from the identification of symmetrical states in such a
way that the relevant behaviour for the considered properties remains
unchanged.

We will now exploit this symmetry to achieve better scalability
in the number of processes for which the model can be effectively
analysed.

The model checker PRISM used in our previous analysis provides a
built-in implementation of a variant of symmetry
reduction~\cite{KNP06} for component symmetry, i.e., where multiple
components are completely indistinguishable from each other. This
approach relies on a canonicalisation of the symmetric states and is
applied directly on the symbolic MTBDD representation of the
model. Unfortunately, the explicit use of the process index in the
spinlock module, necessary to ensure that exactly one of the waiting
processes acquires the lock, prevented an application of this
techniques.

\tudparagraph{Symmetry reduction using generic representatives.}
To apply symmetry reduction, we transformed our model following 
the notion of \emph{generic representatives} as described 
in~\cite{EmTref99,EmWahl03,DM06,DMP09}: 
We pick one arbitrary process, $P_1$, and treat the other $n-1$
processes as indistinguishable. This allowed us to encode the relevant
information in a more succinct form by just counting how
many of the $n-1$ processes are in each of their respective local
states.
As an example, let process $P_2$ be in location
$\mathit{ncrit_2}$ with timer value $t_2=10$, similar to process $P_3$
which is in location $\mathit{ncrit_3}$ with timer value $t_3=10$,
while process $P_4$ is in its location $\mathit{crit_4}$ with
$t_4=2$. The symmetry reduced encoding just records that there
are two processes in a local state with location $\mathit{ncrit}$ and
timer value $t=10$ and that a single process is in the location
$\mathit{crit}$ with timer value $t=2$. It is easy to see that any
other permutation of the process indices $2, 3$ and $4$ for this
configuration would result in the same state in the symmetry-reduced
quotient model. 
The transitions in the quotient model then arise by
combining the effect of the transitions for a given configuration. For
example, in a configuration where all processes are in their
non-critical section with timer values $\geq 0$, a $\mathit{tick}$
transition is enabled that corresponds to all processes staying in
their non-critical section and decreasing their timer value by one. 
In the quotient model, this transition leads to the configuration 
where the number of processes in their non-critical section with timer
value $t$ corresponds to the number of processes in the previous
configuration that are in their non-critical section with timer value
$t+1$. The other possible combinations of state changes in the
processes are handled in a similar fashion.
The symmetry-reduced spinlock model, again a discrete time Markov
chain, then results from the
synchronous product of $P_1$ with the quotient model 
representing the processes $P_2$ to $P_n$ and a slightly adapted 
version of the spinlock process that correctly handles the 
uniform distribution of the lock access between the waiting 
processes by assigning the lock to a waiting process $P_1$ 
with probability $\frac{1}{\mathit{waiting}}$
and with $1 - \frac{1}{\mathit{waiting}}$ to one of the other
processes, where $\mathit{waiting}$ is the total number of processes
currently waiting for the lock.

\tudparagraph{Initial distribution.}
In the non-reduced model, the initial length for the non-critical
section of each process is chosen according to the distribution
$\nu$, i.e., for our example distribution $\nu(40)=\nu(50)=\frac{1}{2}$ each process
starts with probability $\frac{1}{2}$ with a non-critical section of
length $40$ and with probability $\frac{1}{2}$ with a length of $50$. 
In the symmetry-reduced model, the initial configuration of the
symmetric processes $2$ to $n$ is uniformly distributed for the
generic representatives instead, i.e., for 
$\nu(40)=\nu(50)=\frac{1}{2}$ each
assignment of $x$ and $y$ satisfying $x+y = n-1$ is equally likely, 
where $x$ represents the number of symmetric processes starting with
length $40$ and $y$ with length $50$. This modification is benign in
our case, as we are interested in the probabilities in the long run
and every state that occurs infinitely often is reached with
probability $1$ no matter the initial state. For properties that
depend on the precise configuration of the initial states, we can
straightforwardly weigh the initial generic representatives of the
symmetry-reduced model to match the initial distribution of the
original model.

\tudparagraph{Matrix generation for MRMC.} 
We first considered an approach similar to~\cite{DM06,DMP09}, where the
symmetry-reduced model is generated in the PRISM input language. This
approach yielded some improvement, for example for $5$ processes and
$\nu(40)=\nu(50)=\frac{1}{2}$ as in Table~\ref{tab:prism-40-50}, 
the number of reachable states shrank from 198,808,720 to 8,606,543, 
the MTBDD size to 5,192,338, the time for building shrank from
around 20 hours to about 5.5 hours and the time for calculating
the steady-state probabilities was reduced from around 90 minutes to
about 15 minutes. On the other hand, further increasing the number of
processes again resulted in prohibitively long times needed 
to build the model due to the construction of the internal symbolic
representation of the model. 
In addition to using an internal calculation engine, PRISM supports
the export of the transition matrix from its internal representation
of the state space. This, together with the exported state labels,
allows the use of alternative model checkers such as
the probabilistic model checker MRMC~\cite{MRMC11Journal}. 
MRMC operates on a sparse representation of
the transition matrix, i.e., an efficient, compact encoding of
matrices that predominantly consist of entries with the value
zero. The use of MRMC provides a speed-up for calculating the
steady-state probabilities, but the bottleneck of building
the internal symbolic representation by PRISM remains and blocks the
successful scaling of the number of processes in the model.

%

We have thus implemented an approach that instead generates the
reachable state space of the symmetry-reduced model in an explicit
manner and directly outputs the transition matrix and state labels 
of the DTMC in the format usable by MRMC. 
While our tool is tailored to this model, i.e., given the number
of processes $n$ and the parameters $\gamma_0$, $\gamma_1$ and $\nu$
the tool enumerates the reachable state space and calculates the transition
probabilities between the states, the matrix could likewise 
be generated by an explicit state probabilistic model checker like
LiQuor~\cite{CB06b} or even by an explicit state space generator for the PRISM
input language.

\DTLsetseparator{,}
\DTLloaddb[noheader]
{xy}{ssv-includes/data-MRMC-reduced.40.50.csv}

\begin{table}[tbh]
\begin{center}
\begin{tabular}{| r | r | r | r | r |}\hline
  & \headC{Reachable states}& \headC{Time for} & \headC{Time for MRMC} & \headC{Time for} \\
  Processes & \headC{(rows/cols in matrix)} & \headC{matrix generation}  &  \headC{steady state calc.} & \multicolumn{1}{c|}{properties} \\ \hline
  \DTLforeach*{xy}{
     \cola=\dtldefaultkey1,%
     \colb=\dtldefaultkey2,%
     \colc=\dtldefaultkey3,%
     \cold=\dtldefaultkey4,%
     \cole=\dtldefaultkey5,%
     \colf=\dtldefaultkey6,%
     \colg=\dtldefaultkey7%
  }{%
   \ifthenelse{\equal{1}{\DTLcurrentindex}}{}{\\ \hline}%
   \cola & 
   \npthousandsep{,}\numprint{\colb} &
   \npthousandsep{,}\npdecimalsign{.}\nprounddigits{2}\numprint[s]{\cold} &
   \npthousandsep{,}\npdecimalsign{.}\nprounddigits{2}\numprint[s]{\cole} &
   \npthousandsep{,}\npdecimalsign{.}\nprounddigits{2}\numprint[s]{\colf}
  }%
  \\ \hline \hline
  \npthousandsep{,}\numprint{100} &
   \npthousandsep{,}\numprint{205637} &
   \npthousandsep{,}\npdecimalsign{.}\nprounddigits{2}\numprint[s]{2.17} &
   \npthousandsep{,}\npdecimalsign{.}\nprounddigits{2}\numprint[s]{0.79} &
   \npthousandsep{,}\npdecimalsign{.}\nprounddigits{2}\numprint[s]{0.53} \\\hline
  \npthousandsep{,}\numprint{1000} & 
   \npthousandsep{,}\numprint{334337} &
   \npthousandsep{,}\npdecimalsign{.}\nprounddigits{2}\numprint[s]{7.96} &
   \npthousandsep{,}\npdecimalsign{.}\nprounddigits{2}\numprint[s]{1.16} & 
   \npthousandsep{,}\npdecimalsign{.}\nprounddigits{2}\numprint[s]{0.86} \\\hline
  \npthousandsep{,}\numprint{10000} &
   \npthousandsep{,}\numprint{1621337} &
   \npthousandsep{,}\npdecimalsign{.}\nprounddigits{2}\numprint[s]{255.14}   &
   \npthousandsep{,}\npdecimalsign{.}\nprounddigits{2}\numprint[s]{5.18} &
   \npthousandsep{,}\npdecimalsign{.}\nprounddigits{2}\numprint[s]{4.01} \\\hline
\end{tabular}
\vspace{2ex}
\caption{Statistics for directly generating the matrix for the symmetry-reduced model 
  and calculating the steady state probabilities for $n$ processes and
  distributions $\gamma_0(5) = 1$, $\gamma_1(6)=1$ and
  $\nu(40)=\nu(50)=\frac{1}{2}$ 
  using the model checker MRMC.}
\label{tab:mrmc-40-50}
\end{center}
\end{table}

\tudparagraph{Improved scalability.}
Table~\ref{tab:mrmc-40-50} demonstrates the vast improvements in
scalability resulting from our approach of directly generating the
matrix and using MRMC for the calculation of the steady-state
probabilities. The table shows the time spent by our tool to
generate the matrix and the number of reachable states which
were enumerated by our tool. This number corresponds to the number of rows
and columns of the generated matrix. The second-to-last column lists the time
spent by MRMC for calculating the steady-state probabilities for each
state and writing the results to a file. The last column presents 
the time spent in a post-processing step to calculate the steady-state
probabilities for a number of state properties from the output of MRMC,
as detailed below. We have chosen this separate post-processing step
to allow for a more detailed analysis of the results and to facilitate a
simple implementation of the conditional long run operators 
of~\cite{FMICS12} in the future. As our purpose was an initial
evaluation of the potential of symmetry reduction to this 
and similar models, both the matrix generator and the post processing
tool are not yet heavily optimised for speed and space efficiency.

As can be readily seen in Table~\ref{tab:mrmc-40-50}, the complexity
of the model for these parameters 
increases for up to 8 processes, with the model becoming
drastically simpler afterwards. To illustrate this phenomenon and to 
gain further insights in 
the behaviour of the spinlock model 
when the number of processes and the distributions are varied, we have
calculated the steady-state probabilities of a number of state
properties using our tool and MRMC.
 
\begin{figure}[tpbh]
\centering
\includegraphics[width=7.8cm]{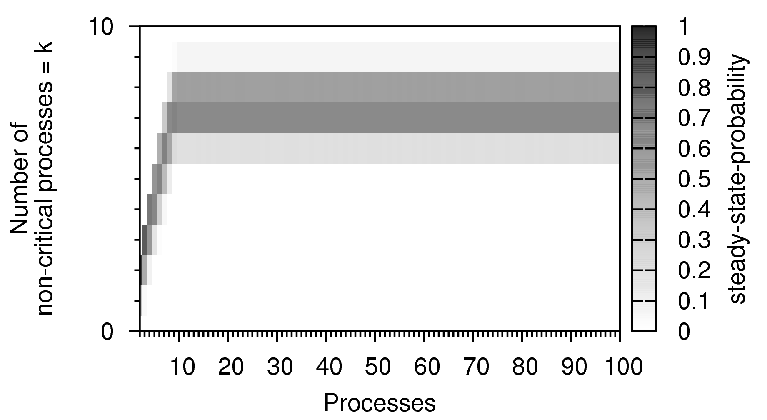}%
\;
\includegraphics[width=7.8cm]{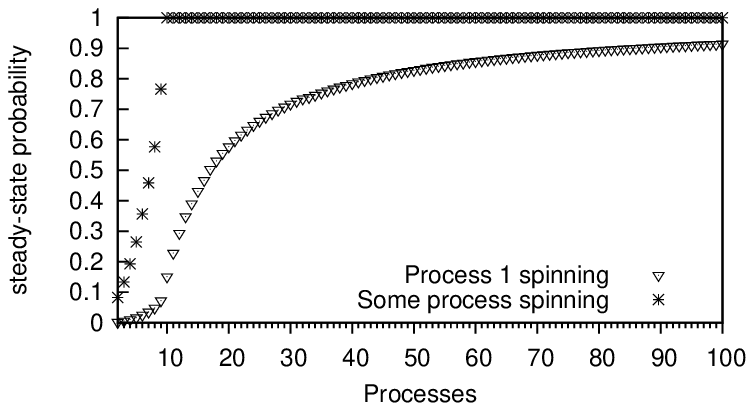}
\caption{The left diagram shows the steady-state probabilities that at a given moment 
         there are exactly $k$ processes in their non-critical
         section. Darker grey indicates higher probability, as
         indicated in the legend.
         The right diagram shows
         the steady-state probabilities that at a given moment 
         (1) process $P_1$ is ``spinning'', i.e., waiting for the lock
         and
         (2) that at least one process is ``spinning''.
         $2$ to $100$ processes.}
\label{fig:sym.5.6.40.50.ncrit.spin}
\end{figure}

Figure~\ref{fig:sym.5.6.40.50.ncrit.spin} shows
the effect of increasing the number of processes 
in our model.
The figure on the left depicts the steady-state probability that at a given
moment exactly $k$ out of the $n$ processes are in their non-critical section for
our example distribution. 
As can be seen, increasing the number of processes at first increases
the number of processes expected to be in their non-critical section because
processes requesting the
lock have a high probability to immediately acquire the lock and thus quickly return to the
non-critical section. At some point the number of processes
reaches a threshold where the lock becomes saturated. At this point, 
the probability that there is already a process waiting for the lock
when the lock is released reaches $1$. This limits the number of
processes that can simultaneously be in their non-critical section to
a fixed number, which is related to the ratio of the longest time a process may
spend in the non-critical section and the shortest time a process
needs to pass through its critical section.
The figure on the right in Figure~\ref{fig:sym.5.6.40.50.ncrit.spin} 
details the effect of lock saturation for an increasing number of
processes. For the chosen values for $\gamma_i$ and $\nu$, saturation
is reached for 10 or more processes, as the probability that in the
long run some process is waiting for the lock at any given moment 
reaches $1$. This entails that each individual process, here represented by
$P_1$, spends a larger and larger amount of its time waiting for the
lock.

\begin{figure}%
\centering
\subfigure[{The steady-state probabilities that at a given moment 
         there are two (or more) processes with distance equal to $k$
         in their non-critical section, 
         for $2$ to $100$ processes and with $\gamma_0(5)=1$, $\gamma_1(6)=1$
         and $\nu(40)=\nu(50)=\frac{1}{2}$. Darker grey
         represents a higher probability as indicated in the
         legend on the right.}]{%
\label{fig:sym.5.6.40.50.dist}%
\includegraphics[width=7.8cm]{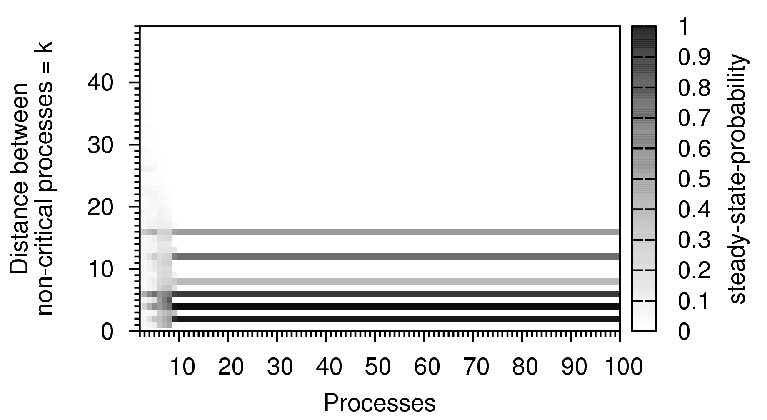}
}
\subfigure[{
  Steady-state probabilities that at a given moment 
  (1) process $P_1$ is ``spinning'', i.e., waiting for the lock
  and
  (2) that at least one process is ``spinning''.
  $2$ to $100$ processes,
  $\gamma_0(5)=1$, $\gamma_1(6)=1$
  and $\nu(50)=\nu(60)=\frac{1}{2}$. The grey bar indicates the
  gap, where the model exceeded the 190GB RAM limit.}]{
\label{fig:sym.5.6.50.60.spin}
\includegraphics[width=7.8cm]{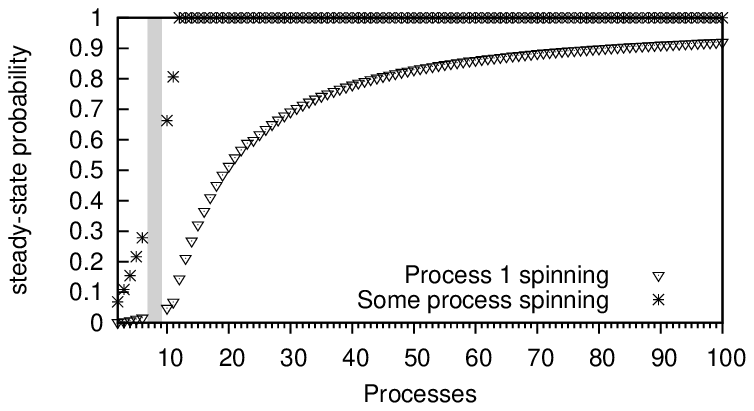}
}
\caption{
}
\end{figure}

Figure~\ref{fig:sym.5.6.40.50.dist} provides a different view on
the same phenomenon. There, we analyse the regularity of the pattern
in which processes pass through their non-critical section. The figure
shows the steady-state probabilities that at a given moment two (or
more) processes in their non-critical section have a distance of $k$
time units, with no other process in between. E.g., two processes in their
non-critical section with timer values of $12$ and $16$ and no process
in between have a distance of $4$. As can be expected, if the number of
processes is sufficiently low to ensure that the lock is not
saturated, the distances between the non-critical processes take a
variety of values as the probabilistic choice of the time spent in the
non-critical section and the two different lengths of the critical
sections ``shuffle'' the order and distance of the
processes. Increasing the number of processes then leads to a much
more regular structure, as the regularity with which processes pass
through the lock induces less variety in the distances between processes.

\setlength{\subfigcapmargin}{5pt}

\tudparagraph{Varying the distribution.} We have seen that our approach
of using symmetry reduction and the model checker MRMC allowed us to
successfully scale the number of processes for the example
distribution $\nu(40)=\nu(50)=\frac{1}{2}$. Unfortunately, 
for other, more challenging distributions considered 
in~\cite{FMICS12}, the
picture is not as rosy. Figure~\ref{fig:sym.5.6.50.60.spin} shows
some results for the distribution $\nu(50)=\nu(60)=\frac{1}{2}$. 
As can be seen, the values for $n=7$ to $9$ processes could not 
be calculated, as the corresponding matrix could not be generated 
before reaching the memory limit of $190$ GB. Similar gaps 
where generating the matrix exceeds the available 
RAM exist also for other distributions we considered.

\tudparagraph{Scalability of saturated locks.} The drastic improvements
in scalability beyond our expectations and the collapse in complexity 
that we observe is linked to the complete or almost complete
saturation of the lock. Increasing the number of processes 
further only increases the number of processes that wait spinning for
the lock.

Of course, spinlocks are known to not scale well in high-contention
scenarios and are therefore not widely used in situations where lock
contention can become high. For practical purposes, parameters that 
lead to lock saturation then strongly indicate that locking realised 
via a spinlock is probably not the right choice and more sophisticated
locking schemes or even architectural redesign of a given system is
required.  However, we expect that the approach presented here 
should apply equally well to other synchronisation primitives 
for highly contended resources and other scenarios such as the
analysis of thread pools where by construction the majority of worker threads await further requests.


\section{Conclusions and Future Work}
\label{sec:conclusions}

In this paper, we have investigated how symmetry reduction helps
scaling a previous case study on model checking conditional long run
properties of low-level operating-system code. Using a model specific
adaption of the technique of generic representatives, we were able to
significantly scale the number of processes in the model beyond the
point were adding more processes does not significantly increase the
complexity of the analysis anymore. This allows an analysis of models
with $10,000$ processes and beyond using the model checker
MRMC. There are still situations where the analysis proves elusive for certain number of
processes due to the high number of reachable states.

\tudparagraph{Future work.}
The results presented in this paper represent work in progress and
allowed us to evaluate the potential benefit of using symmetry reduction
techniques for the scalability in the number of processes 
of a probabilistic analysis of locking mechanisms and similar
low-level constructs. We plan to further improve the time and space
efficiency of the tool we use to generate the state space. We also plan to 
investigate refined notions of symmetry reduction, the compatibility
with symbolic methods, i.e., the encoding of the states and
corresponding variable orders, as well as other reduction techniques
such as bisimulation
quotienting/lumping~\cite{KatoenKZJ_TACAS07,MRMC11Journal}.
As our results show the feasibility of an analysis approach 
using symmetry reduction
and the model checker MRMC, we further plan to extend MRMC by the
capability to calculate conditional long-run properties of the type
considered in~\cite{FMICS12}. 

A second topic for future work are more complicated case
studies. Interesting candidates are thread pools and queue-based
locks~\cite{anderson:spin_lock_alternatives, mcs} as they enforce a
strong dependency between waiting processes. Particularly interesting
due to their potential feedback on the design and use of these
algorithms are reactive locking
schemes~\cite{Lim:1994:RSA:195473.195490}. Probabilistic model
checking results could guide when the lock switches to an alternative
implementation. Overcoming the complexity gap remains a top priority.


\bibliographystyle{eptcs}
\bibliography{ssv}

\end{document}